\documentclass[aps,prb,reprint,superscriptaddress]{revtex4-1}
\usepackage{amsmath}
\usepackage{graphicx}
\usepackage{bm}
\usepackage{amssymb}
\usepackage{hyperref}
\usepackage{xcolor}

\begin{document}

\author{Xi-Rong Chen}
\affiliation{National Laboratory of Solid State Microstructures and department of Physics, Nanjing University, Nanjing, 210093, China}
\affiliation{Collaborative Innovation Center of Advanced Microstructures, Nanjing University, Nanjing 210093, China}

\author{Wei Chen}
\email{Corresponding author: pchenweis@gmail.com}
\affiliation{College of Science, Nanjing University of Aeronautics and Astronautics, Nanjing 210016, China}
\affiliation{Institute for Theoretical Physics, ETH Zurich, 8093 Z\"{u}rich, Switzerland}

\author{L. B. Shao}
\affiliation{National Laboratory of Solid State Microstructures and department of Physics, Nanjing University, Nanjing, 210093, China}
\affiliation{Collaborative Innovation Center of Advanced Microstructures, Nanjing University, Nanjing 210093, China}

\author{D. Y. Xing}
\affiliation{National Laboratory of Solid State Microstructures and department of Physics, Nanjing University, Nanjing, 210093, China}
\affiliation{Collaborative Innovation Center of Advanced Microstructures, Nanjing University, Nanjing 210093, China}

\title{Engineering chiral edge states in 2D topological insulator/ferromagnetic insulator heterostructures}

\begin{abstract}
Chiral edge state (CES) at zero magnetic field has already been realized in the magnetically doped topological insulator (TI).
However, this scheme strongly relies on material breakthroughs, and in fact, most of the TIs cannot be driven into a Chern insulator in this way.
Here, we propose to achieve the CESs  in 2D TI/ferromagnetic insulator/TI sandwiched structures
through spin-selective coupling between the helical edge states of the two TIs. Due to this coupling, the edge states of one spin channel
are gapped and those of the other spin channel remain almost gapless,  so that the helical edge states of each isolated TI are changed into the CESs.
Such CESs can be hopefully achieved in all TI materials, which are immune to magnetic-disorder-induced backscattering.
We propose to implement our scheme in the van der Waals heterostructures between monolayer 1T'-WTe$_2$ and bilayer CrI$_3$.
The electrical control of magnetism in bilayer CrI$_3$ switches the transport direction of the CESs,
which can realize a low-consumption transistor.

\end{abstract}
\maketitle

\section{introduction}
The discovery of quantum Hall effect \cite{Klitzing80prl} has lead to
a revolution in condensed matter physics, that is, understanding phase transition
in terms of band topology \cite{Thouless82PRL}. Following this route,
various topological materials have been found in the last decade, such as
topological insulator (TI), superconductor, and semimetal \cite{Hasan10rmp,Qi11rmp,Armitage18rmp}.
Apart from the significance in fundamental physics,
the chiral edge states (CESs) in quantum Hall phase also has
important applications in both quantum information processing
and dissipationless electron transport \cite{Qi11rmp}. The unidirectional channel of the CESs
avoids all kinds of elastic backscattering, resulting in a
long phase coherence length \cite{Ji03nat,Henny99scn,Neder07nat}
and a low energy dissipation.

The quantum Hall effect is stabilized by a strong magnetic field,
which is inapplicable for a low-consumption integrated circuit
in the future.
Therefore, searching for CESs without magnetic field is of great interest.
Quantum Hall effect without external magnetic field is
called quantum anomalous Hall effect, or Chern insulator, which was first
proposed by Haldane \cite{Haldane88prl}. Recently,
the quantum anomalous Hall effect has been realized
in thin films of chromium-doped (Bi,Sb)$_2$Te$_3$, a
magnetically doped TI \cite{Chang13scn,Yu10scn}.
The spontaneous ferromagnetic order in the TI
breaks time-reversal symmetry, results in a nonzero Chern number.
Accordingly, the helical edge states in
the TI are turned into the chiral ones \cite{Liu08prl,Klinovaja15prb},
the latter being more robust due to the lack of backscattering channels \cite{Li13prl}.
Although the physical scenario of
engineering Chern insulator
by magnetically doping a TI is general,
its realization strongly relies on the details of the bulk material.
In fact, due to the complex interaction between electrons and magnetic
impurities, in most of the cases, a ferromagnetic order cannot
be established in the TI materials.
For example, in HgMnTe, magnetic moments do not order
spontaneously and an additional small Zeeman field is
required \cite{Yu10scn,Liu08prl}.
This brings up an interesting question that
whether or not there is a general way to create CESs
which does not rely too much on specific properties of the bulk
material used.

\begin{figure}
\centering
\includegraphics[width=0.45\textwidth]{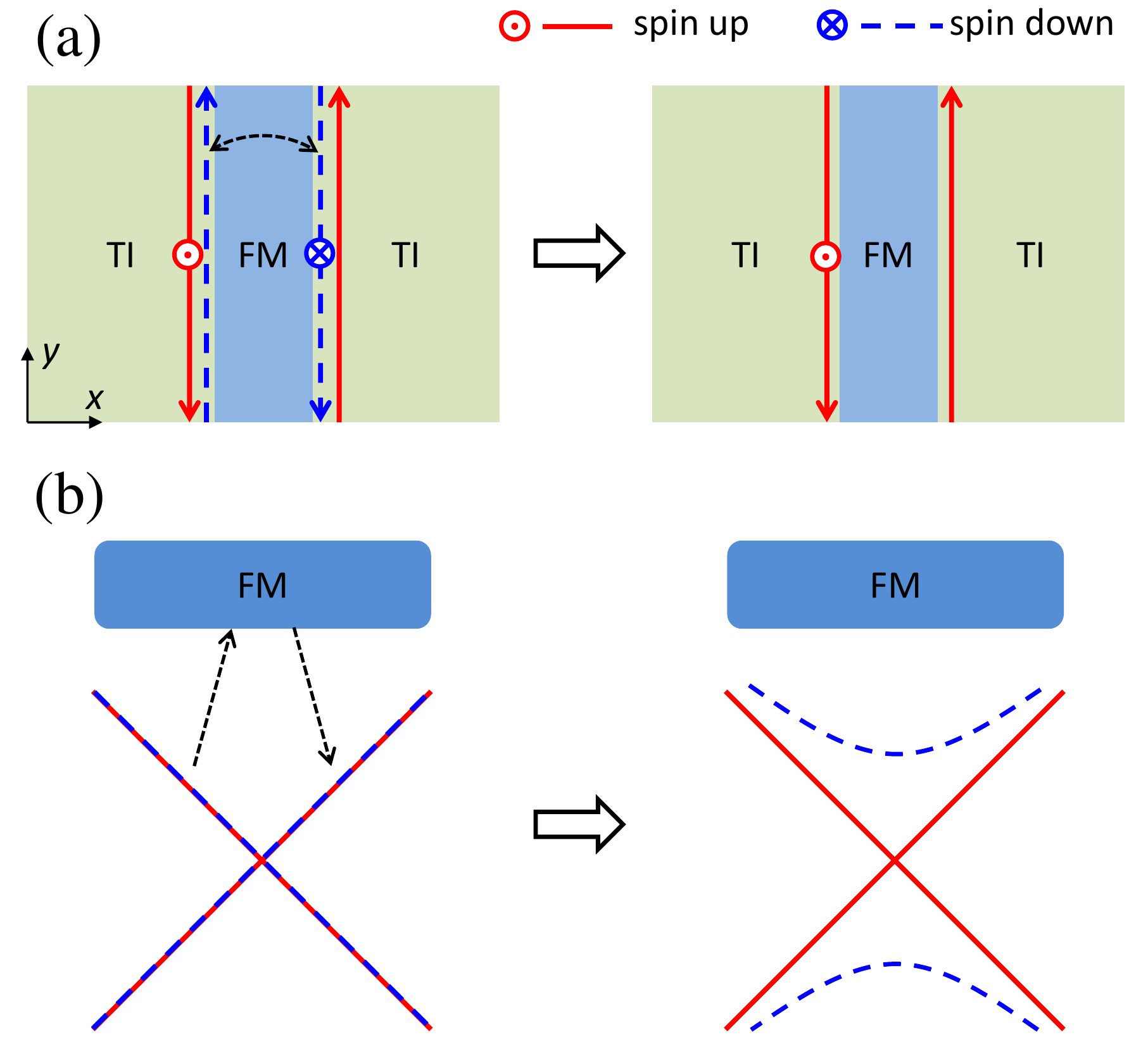}
\caption{(Color online). (a) Illustration of the
TI/FI/TI junction. The FI induces a spin-selective coupling between
the helical edge states in two TIs. A gap is
opened only in the spin-down edge states, leaving
the gapless spin-up edge states to form CESs.
(b)  Modification of the edge-state band
structure by the coupling.}
\label{fig1}
\end{figure}

In this work, we propose to engineer CESs in a 2D TI/ferromagnetic insulator (FI)/TI
sandwiched structure,  which can be hopefully applied to all 2D TI materials.
The Zeeman splitting inside the FI introduces a spin-selective
coupling between two adjacent helical edge states of two TIs
as shown in Fig. \ref{fig1}. As a result of this coupling,
the spin-down edge states are gapped but the spin-up ones remain almost gapless, so that
the helical edge states of each isolated TI are changed into the CESs. The scenario here is
quite different from the magnetically doping method. For the present sandwiched structure, there is no
topological phase transition in the bulk state, and its CESs
are not protected by the bulk band topology. However, such CESs
share all the properties of that in the Chern insulator. Due to the lack of the backscattering channel,
they are very robust against magnetic disorder and favorable to
dissipationless  directed transport. We propose to implement our scheme in the
van der Waals heterostructures
between monolayer 1T'-WTe$_2$ and bilayer CrI$_3$.
The magnetism in bilayer CrI$_3$ can be controlled
by electronic field, which can
switch the transport direction of the CESs,
providing an important application of
a low-consumption transistor.

The rest of this paper is organized as follows. In Sec. \ref{sec2}, we
apply a perturbation approach to the low-energy model of the TI
to show the basic mechanism of our scheme. The perturbation result is further verified by
numerical calculations based on a lattice model in Sec. \ref{sec3}.
In order to show the robustness of the CESs, we calculate
conductance in the presence of magnetic
disorder in Sec. \ref{sec4}. The experimental realization
of our proposal and possible application of the
low-consumption transistor are
discussed in Sec. \ref{sec5}. Finally, a brief summary
is given in
Sec. \ref{sec6}.

\section{basic physical picture}\label{sec2}

The main idea of the CESs engineering is sketched
in Fig. \ref{fig1} in a TI/FI/TI junction,
where an FI film is sandwiched between two TIs along the $x$ direction.
The whole Hamiltonian of this system can be written as
\begin{equation}\label{H}
H=H_{\text{edge}}+H_{F}+H_{T}.
\end{equation}
Here $H_{\text{edge}}$ is the  the low-energy effective
Hamiltonian describing the helical edge states of the two TIs, yielding
\begin{equation}
H_{\text{edge}}=-\sum_{k,\sigma,\tau}v_0k c^\dag_k\sigma_z\tau_zc_k,
\end{equation}
where $v_0$ is the velocity of electrons in the edge channels,
and  momentum $k$ is along the $y$ direction, as seen in Fig. \ref{fig1}(a).
$c_{k}=(c_{k\uparrow L},c_{k\downarrow L},c_{k\uparrow R},
c_{k\downarrow R})^{\text{T}}$ is the Fermi operator with
spin subscripts $\sigma=\uparrow,\downarrow$ and edge ones
$\tau=L,R$, where $\tau=L$ ($R$) denote the left (right) TIs.  The
Pauli matrices $\sigma_{x,y,z}$ and $\tau_{x,y,z}$
operate in spin and left/right TIs, respectively.
The energy spectrum is given by $E_0=\pm v_0k$, which
is measured from the Dirac point at $k=0$. $H_{F}$ is the
Hamiltonian of the middle FI, which can be described by
\begin{eqnarray}
H_{F}&=&\sum_{k} (\varepsilon_{k}+\Delta_{\sigma}) f^\dag_{k\sigma}f_{k\sigma},
\end{eqnarray}
where $f_{k\sigma}$ is the Fermi operator in the FI, $\varepsilon_{k}=\hbar^2k^2/(2m)$,
and $\Delta_{\sigma}$ is the energy difference between the spin-$\sigma$ band bottom and the Dirac point.
The spin dependence of $\Delta_{\sigma}$ arises from the Zeeman spin
splitting in the FI. Without loss of generality,
it is assumed that $\Delta_{\uparrow}$ for the spin-up electrons is
much larger than  $\Delta_{\downarrow}$ for the spin-down electrons.
Moreover, we confine the present study to a single transverse mode in the FI. It is straightforward to extend
the result obtained to the multiple channel case.
The coupling between the FI and the helical edge states of the TI can be captured by
\begin{eqnarray}
H_{T}&=&\sum_{k,\sigma,\tau}t_{k}c^\dag_{k\sigma\tau}f_{k\sigma}+\text{H.c.}
\end{eqnarray}
where $t_{k}$ is the coupling strength between the helical edge
states and the states inside the FI.  Usually, $t_{k}$ is very small compared with $\Delta_\sigma$,
i.e., $t_k\ll\Delta_\sigma$. Taking this fact into account, we
treat the term $H_T$ as a perturbation to $H_{\text{edge}}+H_{F}$ of the decoupled system.

The coupling between the helical edge states and the FI induces a
spin-selective gap opening in the edge states, which can be obtained by solving
the effective model of Eq. \eqref{H}. Note that the spin component
$\sigma_z$ is conserved for the whole system so that the
edge states for each spin polarization
can be solved independently. We formally
write down the Schr\"{o}dinger equation for the spin-$\sigma$ electrons as
\begin{equation}
\left(
  \begin{array}{cc}
    h_\sigma(k) & T(k) \\
    T^\dag(k) & h_{\sigma}^F(k) \\
  \end{array}
\right)
\left(
         \begin{array}{c}
           \psi_\sigma \\
           \psi^F_\sigma \\
         \end{array}
       \right)=E\left(
         \begin{array}{c}
           \psi_\sigma \\
           \psi^F\sigma \\
         \end{array}
       \right),
\end{equation}
where $h_\sigma(k)=-\sigma v_0k\tau_z$, $h^F_\sigma(k)=\varepsilon_{k}+\Delta_\sigma$ and
$T(k)=t_k$, which can be read from Eq. \eqref{H},
and wave functions $\psi_\sigma$ and $\psi^F_\sigma$
refer to the spin-$\sigma$ edge states of the two
TIs and the bulk states in the FI, respectively. By eliminating $\psi^F_\sigma$ in the
two equations above, we obtain the equation for the edge states as
\begin{equation}
\Big[h_\sigma(k)+\Sigma_\sigma(k,\omega)-\omega\Big]\psi_\sigma=0,
\end{equation}
where  $\Sigma_\sigma(k,\omega)=TG_F(\omega)T^\dag$ is  the self-energy with
$G_F(\omega)=[\omega-h^F_\sigma(k)]^{-1}$ as the bare Green's function of the FI.
The self-energy can be solved directly as $\Sigma_\sigma(k,\omega)
=|t_k|^2(\tau_x+\tau_0)/(\omega-\varepsilon_k-\Delta_\sigma)
$, where $\tau_0$ is the identity matrix
in the left/right TI space. We focus on the
parametric region around the Dirac point at which
$k\simeq0$ and $ \omega\simeq0$. Then the self-energy reduces to
$\Sigma_\sigma^0
=-\delta_\sigma(\tau_x+\tau_0)$ with
$\delta_\sigma=|t_0|^2/\Delta_\sigma$.
The effective Hamiltonian for the modified edge states now becomes
\begin{equation}
h'_\sigma(k)=-\sigma v_0k\tau_z-\delta_\sigma\tau_x-\delta_\sigma.
\end{equation}
Correspondingly, the energy spectrum is given by
\begin{equation}\label{e}
E_\sigma(k)=\pm\sqrt{v_0^2k^2+\delta_\sigma^2}-\delta_\sigma.
\end{equation}
One can see that a gap with size $2\delta_\sigma$ appears
at $k=0$, and the band is
shifted by $\delta_\sigma$ as well.
The physical reason is that the hybridization
between TIs and FI introduces an indirect coupling
between edge states in the two TIs and so opens a gap.
The coupling $t_0$ is independent of spin, and spin-dependent
 gap $\delta_\sigma$ is determined by $\Delta_\sigma$.
For $\Delta_{\uparrow}\gg\Delta_{\downarrow}$, corresponding to a strong spin splitting in the FI,
we have $\delta_\uparrow\ll\delta_\downarrow$. In this case,  only the spin-down edge states
are gapped, while the spin-up edge states
remain almost gapless, thus forming CESs, as shown in Fig. 1(b).

From the above perturbation approach, we have seen
that the spin-selective coupling between the two TIs
can drive the original  helical edge states into the CESs.
In the following, we will  perform numerical calculation to verify this conclusion.

\section{lattice model simulation}\label{sec3}
We adopt the Bernevig-Hughes-Zhang (BHZ)
model \cite{Bernevig06scn} of the HgTe/CdTe quantum wells
to describe the 2D TI. It is expected that the main results
do not rely on the specific model, for the
picture based on the edge states in Sec. \ref{sec2} holds generally, independent of specific
TI materials. The BHZ model takes the following form \cite{Bernevig06scn,Konig08jpsj}
\begin{equation}\label{bhz}
H_{\text{BHZ}}=-Dk^2+Ak_x\tilde{\sigma}_x\sigma_{z}-Ak_y\tilde{\sigma}_y+(M-Bk^2)\tilde{\sigma}_z
\end{equation}
where $\tilde{\sigma}_{x,y,z}$ are the Pauli matrices operating on the
$s$ and $p$ orbital degrees of freedom, and $k^2=k^2_x+k_y^2$.
$A, B, D$ and $M$ are the relevant material parameters,
which can be experimentally controlled.
In the long wavelength limit, the BHZ model can be mapped onto a square lattice by
discretizing the continuous model in Eq. \eqref{bhz}.
By the substitution of $k^2=2a^{-2}[2-\cos(k_{x}a)-\cos(k_{y}a)]$,
$k_x=a^{-1}\sin(k_{x}a)$, and $k_y=a^{-1}\sin(k_{y}a)$,
we obtain the tight-binding Hamiltonian as
\begin{eqnarray}
\mathcal{H}_{\text{BHZ}}&=&\sum_{i}c^\dag_{i}H_{ii}c_{i}+\sum_{i}\big(c^\dag_{i}H_{i,i+a_x}c_{i+a_x}+\text{H.c.}\big)\nonumber\\
      &\ & +\sum_{i}\big(c^\dag_{i}H_{i,i+a_y}c_{i+a_y}+\text{H.c.}\big),
\end{eqnarray}
where $c_{i}=(c_{s,i,\uparrow},c_{p,i,\uparrow},c_{s,i,\downarrow},c_{p,i,\downarrow})$
is the Fermi operator on site
$i$ with both orbital and spin components. $i=(i_x,i_y)$ is the index of the discrete
sites in the square lattice, $a_x=(a, 0)$ and $a_y=(0, a)$ are the unit vectors along the
$x$ and $y$ directions, respectively, with $a$ as the lattice constant.
$H_{ii}$ and $H_{i,i+a_x(a_y)}$ are 4$\times$4 block matrices and take
the explicit form
\begin{eqnarray}
H_{ii}&=&-\frac{4D}{a^2}-\frac{4B}{a^2}\tilde{\sigma}_{z}+M\tilde{\sigma}_{z},\nonumber\\
H_{i,i+a_x}&=&\frac{D+B\tilde{\sigma}_{z}}{a^2}+\frac{A\tilde{\sigma}_{x}\sigma_z}{2ia},\nonumber\\
H_{i,i+a_y}&=&\frac{D+B\tilde{\sigma}_{z}}{a^2}+\frac{iA\tilde{\sigma}_{y}}{2a}.
\end{eqnarray}
The band inversion of the HgTe/CdTe quantum wells
occurs at a critical thickness $d_c=6.3$ nm. Here we
take the physical parameters for the system with a thickness
of $7.0$ nm \cite{Konig08jpsj}:
$A=364.5$ nm meV, $B=-686$ nm$^2$ meV,
$D=-512 $ nm$^2$ meV, and $M=-10$ meV. The lattice constant
is set to $a=5$ nm.

The tight-binding model for the FI can be obtained in a similar way, yielding
\begin{equation}
\mathcal{H}_{F}=\sum_{j}f^\dag_{j}\lambda_j f_{j}-\frac{t}{a^2}\sum_{j}\big(f^\dag_{j}f_{j+a_x}+f^\dag_{j}f_{j+a_y}+\text{H.c.}\big),
\end{equation}
where $f_{j}=(f_{j,\uparrow},f_{j,\downarrow})$
denotes the spinful fermion operator in the FI,
$\lambda_i=4t/a^2-\mu+h_z\sigma_z$ includes
chemical potential $\mu$ and Zeeman
splitting $h_z$, and $t$ is the nearest neighbour hopping.
The parameters for the FI are set to $t=1600$ nm$^2$ meV,
$\mu=-483.5$ meV, and $h_{z}=500$ meV.
Due to the Zeeman exchange field, the spin-up
and spin-down bands split. In the following,
we adopt multiple bands with slight splitting for the FI
to simulate a reasonable bulk density of states.

\begin{figure}
\centering
\includegraphics[width=0.5\textwidth]{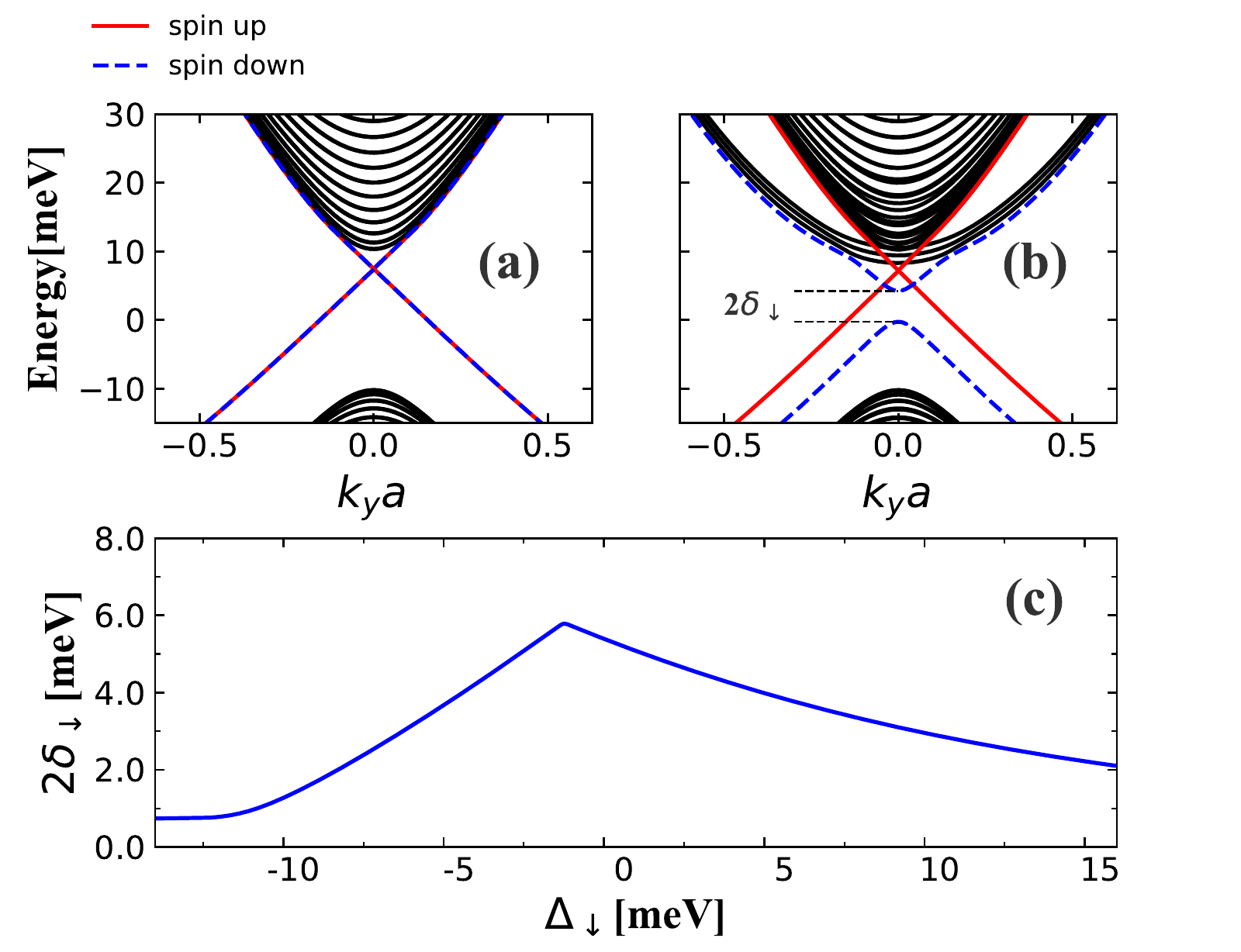}
\caption{(Color online). (a) The band structure of two separated TIs.
(b) The band structure of the TI/FI/TI
junction with the coupling between TI and FI.
The red solid line and blue dashed line represent
spin-up and spin-down edge states, respectively. The
black parabolas in (b) in addition to (a) are the multiple bands in FI.
(c) The gap $2\delta_\downarrow$ for the spin-down edge states
as a function of $\Delta_\downarrow$.
The widths of the TI and FI in the $x$ direction
are 500 nm and 20 nm, respectively.}
\label{fig2}
\end{figure}

The whole hybridized system has a strip geometry along the
$y$ direction, see Fig. \ref{fig1}(a). Since we are only
interested in the edge states at the TI/FI interfaces, in the
numerical calculation, the outer boundaries of two TIs are assumed to
connected with each other by the same hopping strength as in the bulk TI.
Therefore, the system is equivalent to a cylinder. The band structure
is calculated using the Kwant program \cite{Groth14njp}, and the calculated result
is shown in Fig. \ref{fig2}.
In the absence of interface coupling
between the TIs and  FI, two TIs are isolated, and each one contains
a pair of gapless helical states at the interface, see Fig. \ref{fig2}(a).
As the interface hopping is turned on with a strength of $t'=375$ nm$^2$ meV,
a large energy gap is induced for the spin-down channel, but the spin-up channel
remains almost decoupled and gapless, as seen in Fig. \ref{fig2}(b).
These results are in good agreement with the analytic result obtained in Sec. \ref{sec2}.
In addition of the gap opening, the band is shifted in Fig. \ref{fig2}(b), which
was also predicted in the analytic result above.
We also investigate the effect
of the chemical potential $\mu$ in the FI.
Tuning $\mu$ results in a change of the energy gap $\Delta_\downarrow$ in the FI.
The induced gap $2\delta_{\downarrow}$ of the spin-down
edge states as a function of $\Delta_\downarrow$ is shown in Fig, \ref{fig2}(c).
For a large $\Delta_\downarrow$,
they are approximately related by $\delta_\downarrow\propto1/\Delta_\downarrow$
based on the perturbation result in Sec. \ref{sec2}.
However, as $\Delta_\downarrow$ decreases to a small value,
$\delta_\downarrow$ decays as well, because in that case
the minimal gap for the CESs is determined
by $\Delta_\downarrow$ in the FI.
It should be noted that for real materials,
more effects of magnetic proximity
should be taken into consideration, including the
redistribution of charge and spin at the interface,
and band bending effect, etc. \cite{Eremeev13prb,Menshov13prb,Eremeev15JMMM}.

\section{conductance calculation}\label{sec4}

In the 2D TIs, the helical edge states are protected by
time-reversal symmetry, which can be regarded
as two copies of CESs forming
a Kramers pair \cite{Kane05prl}.  For impurity potential
that obeys time-reversal symmetry, helical electrons
cannot be backscattered in the absence of interactions \cite{Xu06prb},
because any backscattering
requires a spin flipping due to the spin-momentum locking.
However, a static magnetic impurity can induce
spin-flip backscattering in the helical edge states and lead to a
suppression of the conductance \cite{Qi11rmp}.
In  the 2D TIs, perfect quantization of the edge conductance has not been achieved so far
\cite{Konig07scn,Roth09scn,Knez14prl,Du15prl,Wu18scn},
in contrast to
the high-quality quantum Hall plateaus \cite{Klitzing80prl,Chang13scn}.
The CESs are very robust
against impurity and disorder, no matter whether they are magnetic or non-magnetic,
because of the absence of backscattering channel.
In what follows we numerically investigate
the effects of magnetic disorder on the
conductance in both helical and chiral edge
states. It is expected that
the CESs achieved by band
engineering in Fig. \ref{fig1} are more robust
against magnetic disorder, resulting in
dissipationless transport.

The lattice model of the system is the same as that in
Sec. \ref{sec3}. In addition, we add magnetic
disorder to the whole system, whose Hamiltonion is given by
\begin{equation}
\mathcal{H}_{\text{dis}}=\sum_iw_ic_i^\dag\sigma_x c_i+\sum_jw_jf_j^\dag\sigma_x f_j,
\end{equation}
where $w_{i,j}$ are the random on-site spin flipping.
We adopt an uncorrelated Gaussian distribution of $w_{i,j}$ with
strength $W$. The transport occurs along the $y$ direction
and the differential conductance
is calculated using the KWANT program \cite{Kwant}.

\begin{figure}
\centering
\includegraphics[width=0.5\textwidth]{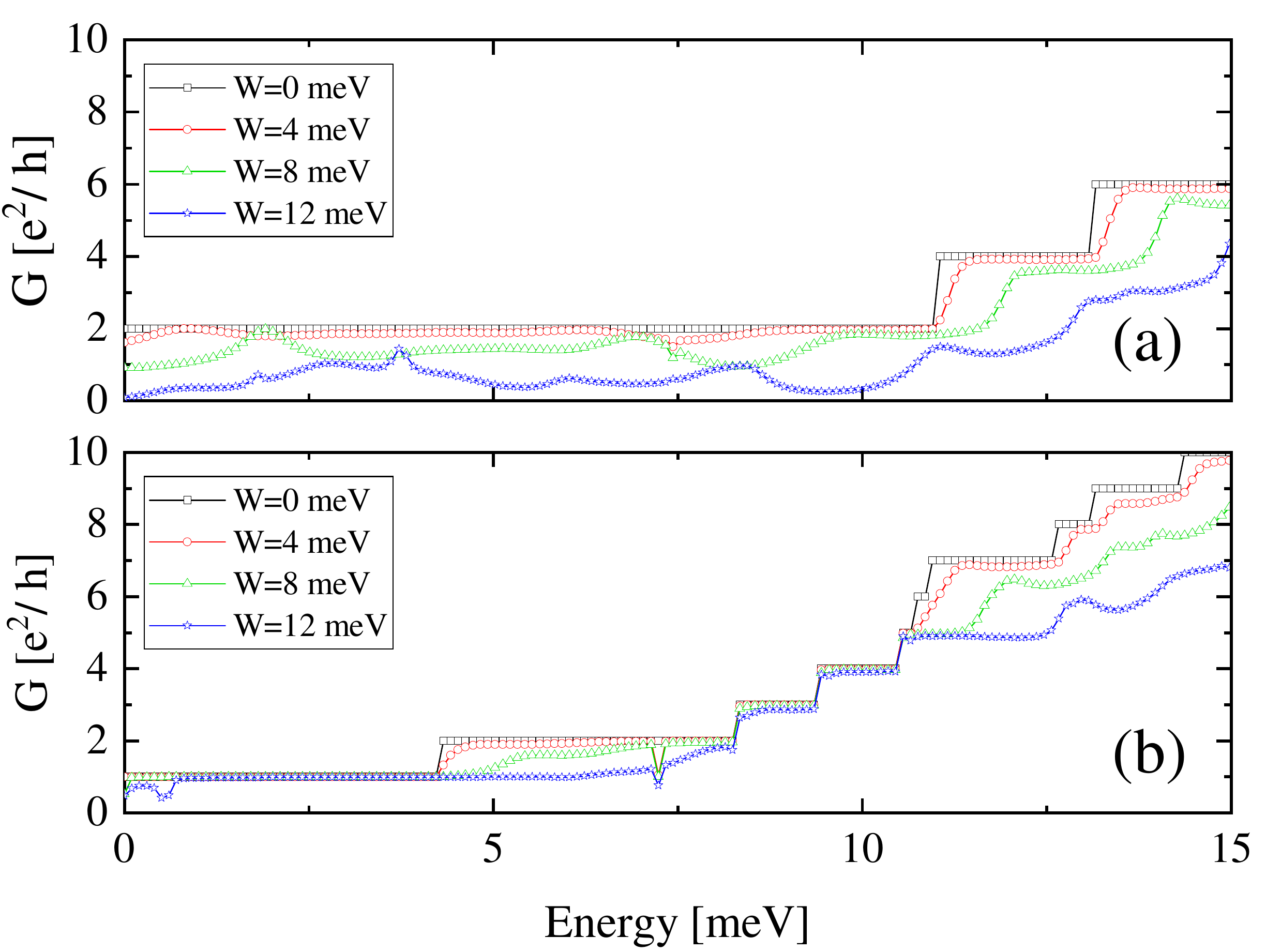}
\caption{(Color online). The conductance of the (a) helical and
(b) chiral edge states as a function of the
incident energy under different magnetic disorder strengths $W$.
The size of the disordered region
in the $y$ direction is 400 nm.}
\label{fig3}
\end{figure}

\begin{figure}
\centering
\includegraphics[width=0.5\textwidth]{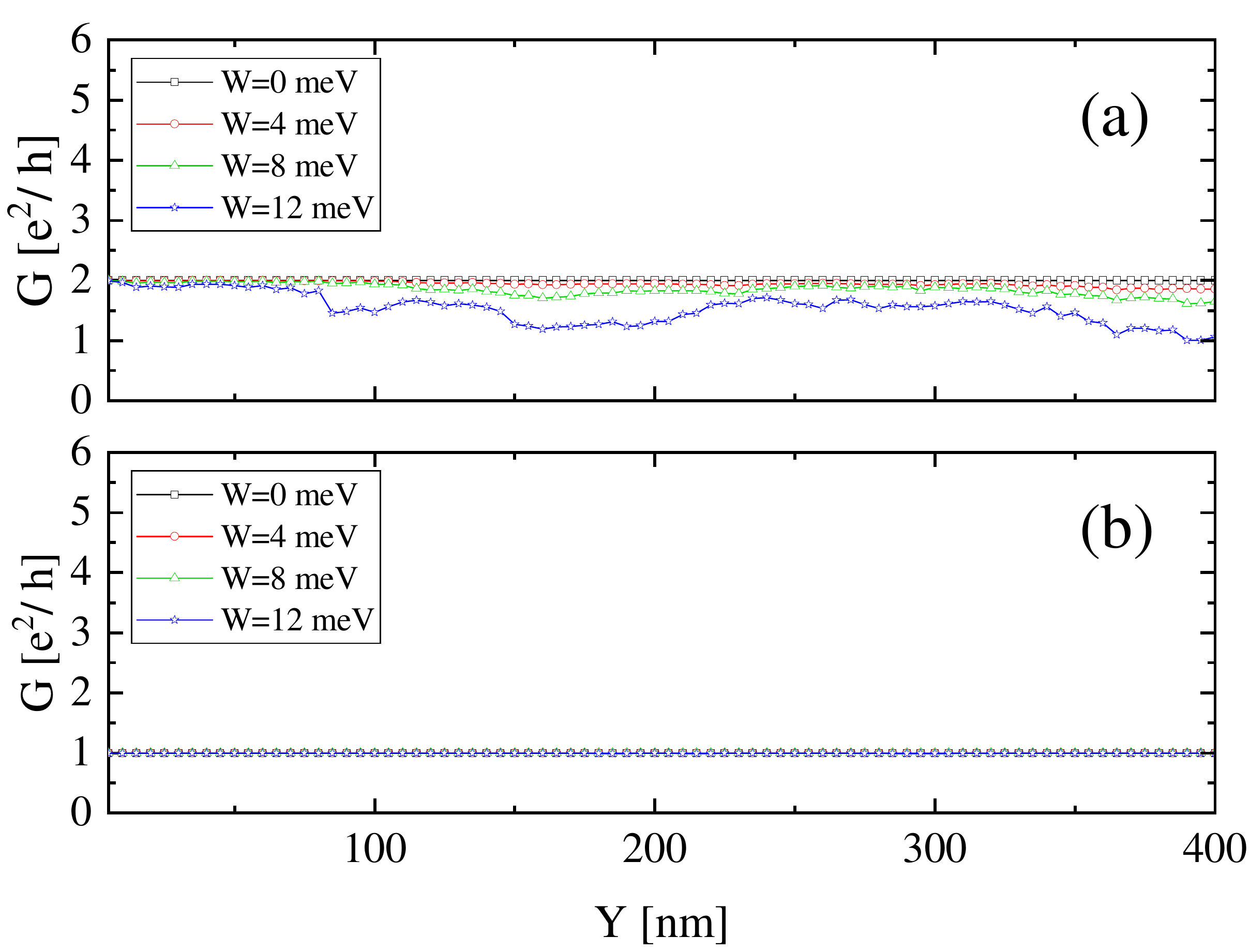}
\caption{(Color online).
The conductance of the (a) helical and
(b) chiral edge states as a function of the length $Y$ of the
disordered region in the $y$ direction. The incident energy is set to 2.5 meV.}
\label{fig4}
\end{figure}

\begin{figure}
\centering
\includegraphics[width=0.45\textwidth]{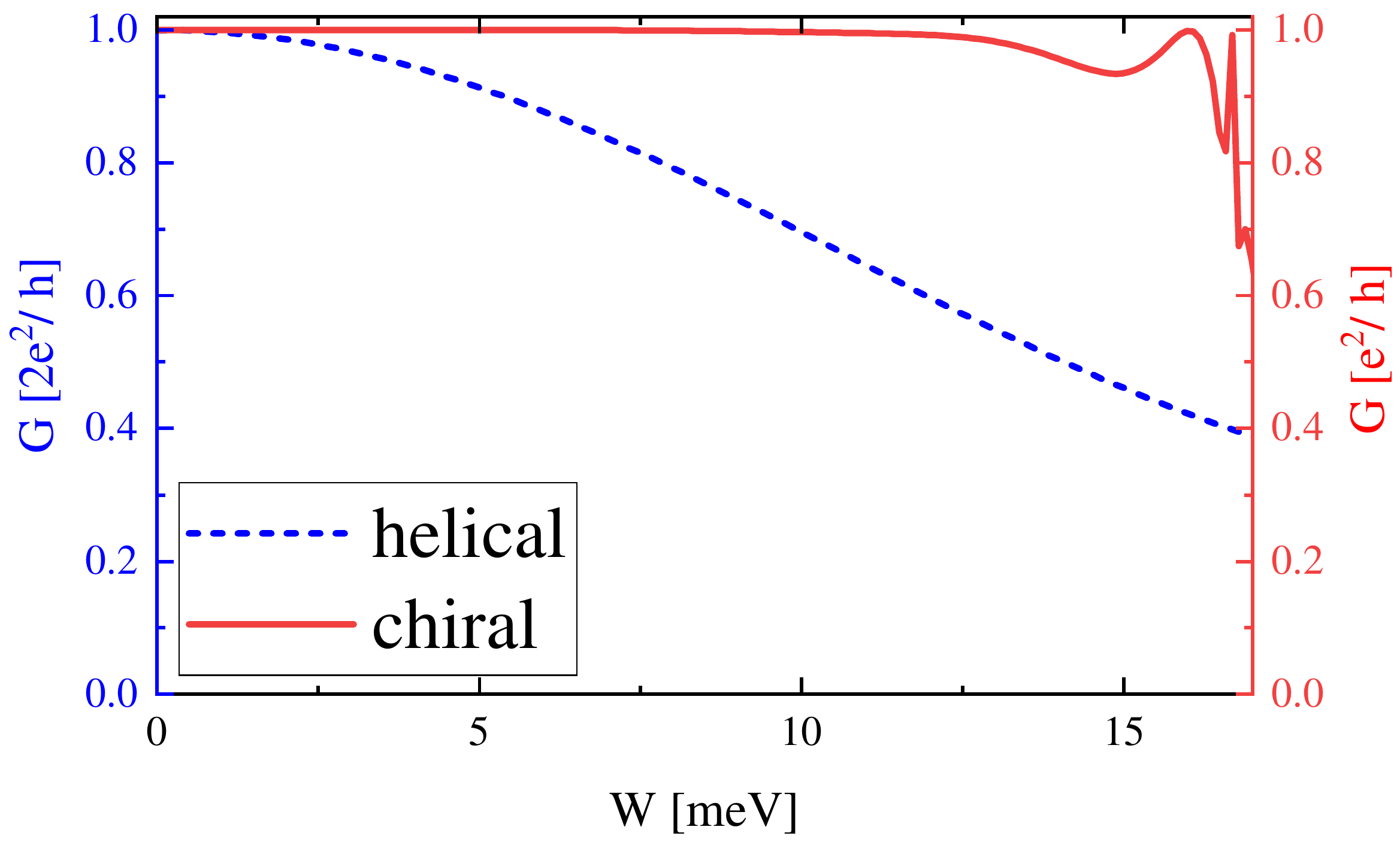}
\caption{(Color online).
The conductance of the helical and chiral edge
states as a function of disorder strength.
The incident energy is set to 2.5 meV and the
length of the disordered region is 400 nm.}
\label{fig5}
\end{figure}

We compare the calculated conductances of the chiral
and helical edge states in
Figs. \ref{fig3} and \ref{fig4}, which
can be achieved by turning on and off the interface hopping
between TIs and FI in the calculation.
Figure \ref{fig3} shows the conductance as a function of incident
energy for different strengths of magnetic disorder.
Without disorder, the helical and chiral edge
channels show quantized conductances $2e^2/h$ and $e^2/h$, respectively.
The lowest plateau is contributed by the edge states.
For the helical edge states in Fig. \ref{fig3}(a),
the conductance is sensitive to the magnetic disorder.
For $W=10$ meV, the conductance
is strongly suppressed. In contrast,
the effect of magnetic disorder on the
CESs is much weaker, and the conductance
remains quantized within the energy gap
induced by the spin-selective coupling [cf. Fig. \ref{fig2}(b)].
The inter-edge reflection (between two TIs) of the CESs induced by magnetic impurities
through virtual scattering between
different spin states can be evaluated as
$\alpha W^2/\delta_\downarrow$, where
$\alpha$ is the spatial overlap
between the spin-up and spin-down edge states.
Since the spin-down states in the two TIs penetrate into
the FI and couple with each other while the spin-up
states are still well localized,
the overlap of their wave functions is small,
indicating a weak backscattering \cite{Li13prl}.
For a fixed incident energy, the
conductance as a function of the length
$Y$ of the disordered region in the $y$ direction
is shown in Fig. \ref{fig4}.
Similar to the results in Fig. \ref{fig3},
in the presence of magnetic disorder, the conductance
for the helical edge states
decreases with the length of the system, indicating
a strong dissipation induced by the magnetic
disorder. However, the quantized conductance through
the CESs retains
with increasing size of the system.
We further compare
the dependence of edge conductance
on disorder strength $W$ in Fig. \ref{fig5}.
One can see that the conductance of
helical edge states decrease much
faster than that of the CESs,
which shows the robustness of the CESs via band engineering.
Up to the disorder strength $W=12$ meV, which is
much larger than the energy gap $\delta_\downarrow$ of the CESs,
the transport of the CESs is always dissipationless.

\section{experimental realization}\label{sec5}

\begin{figure}
\centering
\includegraphics[width=0.45\textwidth]{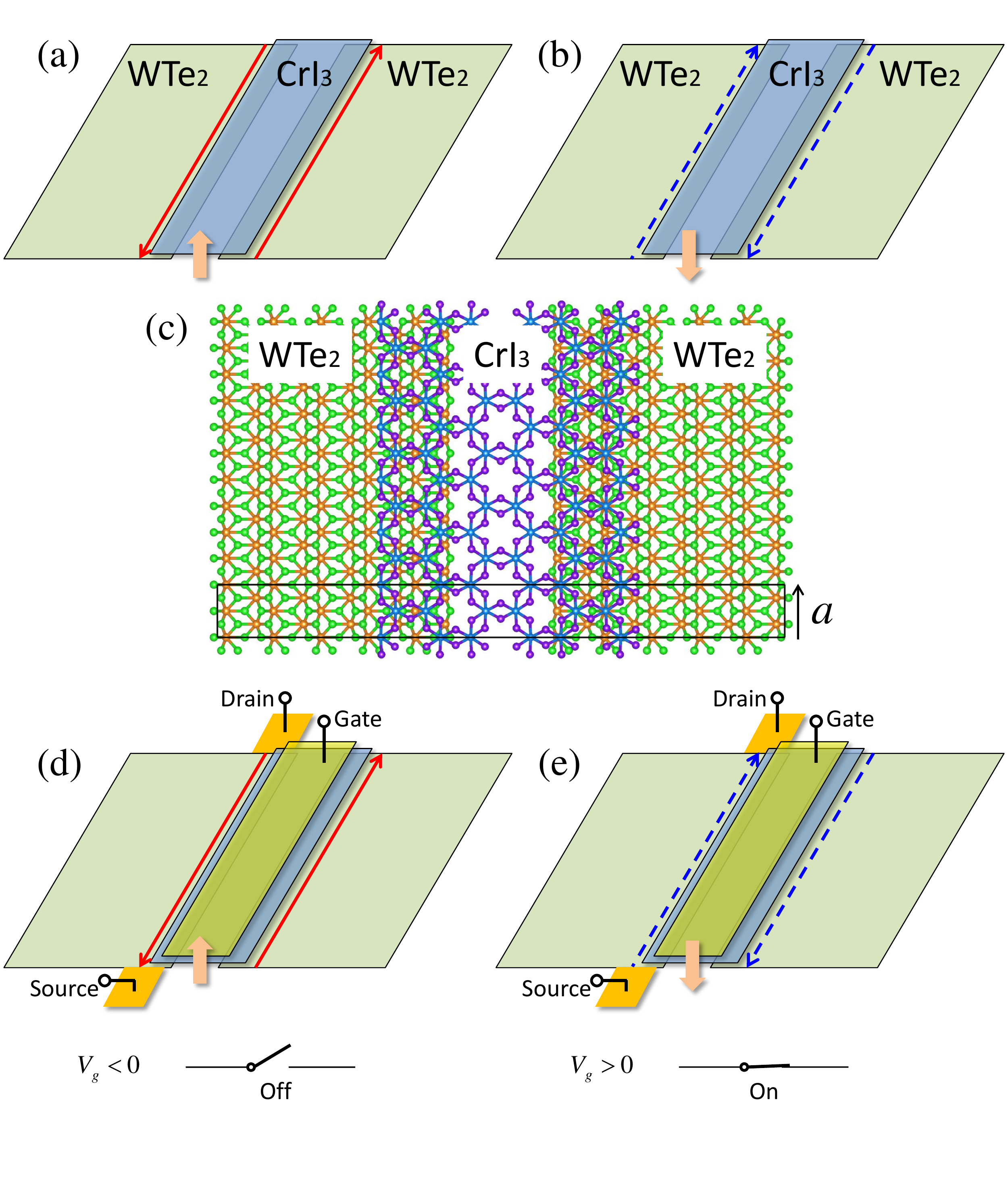}
\caption{(Color online). (a) and (b) Sketch of the van der Waals
heterostructures of monolayer 1T-WTe$_2$ and bilayer
CrI$_3$ with opposite directions of the
Zeeman exchange field. (c) Crystal structure of the system.
The black box denotes the unit cell in the $a$-direction. (d) and (e) Schematic of
the CES-based transistor. The magnetism of
CrI$_3$ is tuned by the gate, which switches on and off
the transport channel between the source and the drain.
}
\label{fig6}
\end{figure}

We would like to discuss
the experimental realization of our proposal.
The building blocks for the CESs in
our proposal are
the 2D TI and FI. Many materials
of TI have been reported \cite{Ando13jpsj}, such as
quantum wells \cite{Konig07scn,Roth09scn,Knez14prl,Du15prl},
the thin film of 3D TIs \cite{Liu10prb}
and recently discovered single-layer transition
metal dichalcogenides \cite{Qian14scn,Wu18scn,Peng17nc,Tang17np}.
Moreover, the TI/FI heterostructures have also been synthesized \cite{Fan14natmat,
Kou13nano,Mellnik14nat,Wang16prl,Fan16natnano,Baker15sr},
so that our proposal can be hopefully realized
by using the state-of-the-art technique in this research field.
In order to achieve the CES engineering,
there are several conditions to be fulfilled. First,
the FI needs to be an insulator, otherwise
the edge states will merge into the bulk states in
the FI. Second,
to achieve a spin-selective coupling between
the helical edge states,
one of the spin states needs to
penetrate into the FI region.
Therefore, the energy difference between the band bottom of the FI and
the Dirac point in the helical edge states should have
a proper value, and the width of the FI in the $x$ direction
should be comparable with the spreading of
the edge states.
Third, a big Zeeman filed is favorable for the
band engineering, which guarantees a negligible
coupling for the edge states with opposite spins.

Recent progress on the 2D van der Waals crystals
has shown that both TI and FI can be
achieved in monolayer or bilayer
crystals \cite{Qian14scn,Wu18scn,Peng17nc,Tang17np,
Gong17nat,Huang17nat,Samarth17nat,Huang18natnano}.
The benefit
of these materials is that
they can be reassembled into designer heterostructures
made layer by layer in a precisely chosen sequence \cite{Geim13nat}.
Moreover, the physical parameters in
such 2D materials can be easily tuned
by gate voltage, such as the chemical potential
and even magnetism \cite{Huang18natnano}.
Thus these materials open a new avenue
to realize our proposal in the van der Waals heterostructures.
We propose to use monolayer 1T'-WTe$_2$ as a
TI \cite{Qian14scn,Wu18scn,Peng17nc,Tang17np} and
the bilayer CrI$_3$ as an FI \cite{Huang18natnano}
to synthesize the heterostructures, see
Fig. \ref{fig6}.
Based on the lattice parameters for both materials \cite{Zheng16am,Mcguire15cm},
the monolayer 1T'-WTe$_2$ and bilayer CrI$_3$ lattices can match
very well along the $a$-direction as shown in Fig. \ref{fig6}(c).
The helical edge states can be
obtained in the 1T'-WTe$_2$ nanoribbon
periodic along the $a$-direction \cite{Zheng16am},
which can serve as a good candidate for the CESs engineering.
1T'-WTe$_2$ has a large topological gap $\sim0.1$ eV \cite{Qian14scn}
which results in the quantum spin Hall effect up to 100 K \cite{Wu18scn}.
The large bulk gap of the TI can also support
a large gap opening in the CESs (Fig. \ref{fig2}),
since  the gap opening in the edge states
is limited by the bulk gap of the TI.
The advantage of CrI$_3$ is that
its magnetism can be electrically controlled \cite{Huang18natnano},
which means that the intrinsic Zeeman field
can be tuned by an electric gate \cite{Matsukura15natnano}.
This remarkable effect opens the possibility
to achieve a low-consumption transistor based
on the electrically tunable CESs, see Figs. \ref{fig6}(d), \ref{fig6}(e).
The direction of the ferromagnetic order in the FI
is controlled by gate voltage $V_g$
and then it determines
the spin and transport direction of the CESs.
For example, when $V_g<0$ there is no CESs
flows from the source to the drain so that
the transistor is turned off [Fig. \ref{fig6}(d)].
Oppositely, when $V_g>0$ the CESs
flow in the opposite direction and the
transistor is turned on [Fig. \ref{fig6}(e)].
The robustness of the CESs
guarantees a very low power dissipation
and strongly suppresses the heat generation.
By stacking multiple layers of the system,
which is the main advantage of the van
der Waals heterostructures,
the on/off ratio of the transistor
can be enhanced \cite{Qian14scn,Liu14natmat}.
As a result, a dissipationless transistor
based on the CESs engineering
can be hopefully implemented.

\section{summary}\label{sec6}
To summarize, we have shown that
CESs can be engineered in
the TI/FI/TI junction. The middle FI film can introduce
a spin-selective coupling between the helical edge states
in the TIs and so open a gap in the edge states of one spin channel.
The edge states of the other spin channel remain almost gapless,
and the resulting CESs
are much  more robust against magnetic disorder
than the helical edge states. We proposed to
implement such CESs in
the van der Waals heterostructures of
monolayer 1T'-WTe$_2$ and bilayer CrI$_3$.
The recently achieved electric control of
magnetism in bilayer CrI$_3$ indicates
that a CESs based low-consumption transistor
can be realized by our proposal.

\begin{acknowledgments}
We thank J. L. Lado, O. Zilberberg and Yong-Ping Du for
helpful discussions.
W.~C. acknowledges the support from the National Natural Science Foundation of
China under Grants No. 11504171.
D.~Y.~X. acknowledges the support from the State Key Program for
Basic Researches of China under Grants No. 2017YFA0303203.
\end{acknowledgments}


%

\end{document}